\documentclass[preprint, notrackchanges]{aastex631} 

\usepackage{multirow, url, color}
\usepackage{bbding, amssymb} 
\usepackage{ulem} 
\usepackage{verbatim} 
\usepackage{enumerate}

\usepackage{amsmath} 
\usepackage{booktabs} 
\usepackage{graphicx, subfigure}
\usepackage{amssymb}
\usepackage[fleqn]{mathtools}
\usepackage{mathrsfs}
\colorlet{myred}{black} 
\begin{document}
\title{
\textcolor{myred}
{An Adaptive Approach for Modeling Collisionless Plasma Shocks and Ion Heating in the Shock Frame}
}
\author[0000-0002-6362-3112]{Yuri A. Omelchenko}
\affiliation{Space Science Institute, Boulder, CO}
\author[0000-0002-6118-0469]{Igor V. Sokolov}
\author[0000-0003-3936-5288]{Lulu Zhao}
\author[0009-0002-4635-3824]{Keheng Zhu}
\affiliation{Department of Climate and Space Sciences and Engineering, University of Michigan, Ann Arbor, MI 48109, USA}
\author[0009-0004-2719-2425]{Michal Golan}
\affiliation{Department of Physics, Ben-Gurion University of the Negev, Beer-Sheva 8410501, Israel}
\begin{abstract}
We present a novel computational method for hybrid (kinetic ions, quasi-neutral fluid electrons) simulations of non-relativistic magnetized collisionless plasma shocks, observed near the Sun, in the solar system, and beyond. 
\textcolor{myred}
{This Adaptive Frame-Of-Reference Approach (AFORA) enables simulations of shocks in their co-moving frames of reference, where shock evolution remains quasi-steady for quasi-stationary upstream conditions, in contrast to moving shocks. We demonstrate our method in two spatial dimensions for a range of angles between the background magnetic field and the shock normal (obliquities). Shock-frame simulations, conducted with a time-adaptive hybrid code, HYPERS, are shown to self-consistently capture upstream and downstream magnetic fluctuations commonly observed by spacecraft at interplanetary shocks and Earth's bow shock. These fluctuations, which may be crucial for estimating the efficiency of ion acceleration at heliospheric and astrophysical shocks, are notably missing from many hybrid simulations of shocks, unless upstream external turbulence is imposed. The quasi-stationary behavior of numerical shocks in the shock frame is confirmed in a series of animations and by demonstrating converged ion energy spectra in the downstream region. Our approach underscores computationally efficient and physically more accurate modeling of shock-driven turbulence, downstream plasma relaxation, and concomitant ion heating in the shock frame. The new method can be used for exploring ion relaxation at interplanetary (IP) shocks, more efficient studies of cosmic rays (CRs) and plasma turbulence, and more accurate predictions of seed populations of Solar Energetic Particles (SEPs) at heliospheric shocks driven by coronal mass ejections (CMEs). 
}
\end{abstract}
\section{Introduction}
\label{sec:intro}

Collisionless plasma shocks occur frequently across the Universe. For instance, solar coronal shocks are created by solar outbursts across all scales, and interplanetary (IP) shocks are associated with Coronal Mass Ejections (CMEs) and corotating interaction regions in the background solar wind. Collisionless plasma shocks also develop when supermagnetosonic plasma flows hit magnetized obstacles such as Earth, or when objects such as planets and galaxies move quickly through the interstellar medium. Importantly, properties of heliospheric and IP shocks can be measured in observations and directly compared to simulations and theory \citep{sokolov2004new, liu2025physics, gedalin2023, gedalin2026}. Supercritical plasma shocks are known for their ubiquitous ability to efficiently convert energy of the directed plasma flow into energy of energized charged particles and magnetic field fluctuations \citep{balogh2013}. 

This work focuses on modeling non-relativistic magnetized collisionless shocks. When a shock generally propagates in a plasma immersed in a magnetic field, ${\bf B}_0$, the normal vector to its front surface, $\bf n$, forms an angle, $\theta_{Bn}$, with respect to ${\bf B}_0$ (further referred to as shock obliquity). Mechanisms for ion acceleration at quasi-parallel ($\theta_{Bn}<45^\circ$) and quasi-perpendicular ($\theta_{Bn}>45^\circ$) shocks have been the subject of extensive observational, theoretical, and computational studies \citep[e.g., see][]{balogh2013}. In this paper, we characterize simulation shocks using the supplementary angle, $\theta_{B}=180^\circ-\theta_{Bn}$, that the ambient magnetic field, ${\bf B}_0$, forms with the solar wind velocity, ${\bf V}_{SW}$, anti-parallel to the shock normal, $\bf n$, in the simulation frame. In our analysis, $\theta_B$ serves the same purpose as $\theta_{Bn}$ in observations.

To better understand physical mechanisms for particle energization by collisionless plasma shocks, one needs to carefully study spatiotemporal ion-scale processes in the shock upstream, transition, and downstream regions. For instance, evaluation of the efficiency of ion acceleration at CME-driven shocks requires a self-consistent analysis of interactions of ions (generally protons and heavy ions) with a moving shock on the ion inertial scales, followed by long-time acceleration and transport of ions by solar wind turbulence. These effects are believed to result in the establishment of energetic ion distributions of Solar Energetic Particles (SEPs), measured by spacecraft at different distances from the Sun \citep{balogh2013}. 

The acceleration of shock-accelerated (``seed'') ions over large spatial scales is modeled by Fokker-Planck (``diffusion'') codes \citep[e.g., see][]{sokolov2004new}, which advance an initial velocity distribution of solar ions in space and time during their transport through the background plasma turbulence. For self-consistent studies, one needs to solve one of the most critical problems in shock physics, known as the ``injection problem'' \citep{balogh2013}. Essentially, this problem can be reduced to finding the number density of seed ions with energies exceeding a certain ``injection energy'', needed for starting the long-time ion energization cycle by plasma turbulence \citep{sokolov2004new}. 

Self-consistent simulations of shock-driven particle acceleration \citep{giacalone2005, giacalone2020solar, gargate2012, gedalin2023} utilize hybrid plasma models, which treat ion components of plasma as fully kinetic species (e.g., particles) and the plasma electrons as an inertia-free fluid in the quasi-neutral Darwin approximation \citep{Winske2003}. Because ion acceleration on their inertial scales is a slow process, multi-dimensional hybrid-PIC (Particle-in-Cell) simulations require immense computational times and simulation domains, with high-resolution three-dimensional (3D) simulations still remaining out of reach even on modern supercomputers. 

\textcolor{myred}{In astrophysical studies of the problem of cosmic ray (CR) acceleration, the precise mechanisms by which CR ions are accelerated at shocks and their
maximum energies are still unknown. These uncertainties remain in hybrid simulations largely because energetic ions constantly modify the shock structure and drive plasma oscillations upstream of a shock  \citep{simon2026}. Most of these simulations explore moving shocks initialized by the ``piston method'', where the plasma flow hits a reflective wall so that the reflected flow collides with the incoming flow. While the shock propagates through the fixed simulation domain, its upstream region is gradually shrinking in time, which limits the available space for instabilities to grow, thus limiting the ability of the simulation to predict the efficiency of CR acceleration \citep{bennett1995, giacalone1997, simon2026}.
}

\textcolor{myred}{The ``piston method'' does not mimic actual astrophysical or solar physics scenarios, where shocks, generated by powerful sources, propagate through their ambient plasmas. In fact, this numerical method falls short of maintaining a physical boundary condition at the reflecting wall. A shock must propagate far enough from this boundary so that this boundary does not interfere with the self-consistent processes taking place near the shock front and downstream of it. The wall effects can clearly be observed in unique animations created in this study \citep{Shockmovies2026}.
}

\textcolor{myred}{Below, we describe an alternative self-consistent technique for modeling the dynamics of a collisionless plasma shock and concomitant ion acceleration in the ``shock frame'', i.e., the frame co-moving with the shock front and characterized with ``free'' downstream boundary conditions. We show that compared to the ``piston method'', our approach forms the more physical shock front, upstream, and downstream regions, while
also achieving obvious performance gains, discussed by \citet{bennett1995} and by \citet{simon2026}. We show that our self-consistent approach offers computational and physical advantages for studying ion acceleration in the downstream and upstream regions of heliospheric, interplanetary (IP), and astrophysical shocks. 
}

In Section~\ref{sec:limits}, we review current limitations of kinetic (full-PIC and hybrid-PIC) models of collisionless plasma shocks. In Section~\ref{sec:AFORA}, we introduce a novel, Adaptive Frame-Of-Reference  Algorithm (AFORA) for hybrid simulations of collisionless plasma shocks and concomitant ion acceleration. In Section~\ref{sec:HYPERS}, we apply this algorithm to 2D shock simulations, using an asynchronous, event-driven hybrid code, HYPERS \citep{Omelchenko2012a, Omelchenko2023spr}. In Section~\ref{sec:disc}, we compare shock profiles from our simulations to observations and other hybrid simulations. Finally, in Section~\ref{sec:summ}, we summarize the most important findings of this study.

\section{Overview of Kinetic Simulations of Collisionless Shocks} \label{sec:limits}

\subsection{Moving Shock Simulations}

\textcolor{myred}{Hybrid-PIC models, which account for ion inertial and cyclotron scales, need to resolve the ion inertial length, $d_i=c/\omega_{pi}$, where $\omega_{pi}$ is the upstream ion plasma frequency and $c$ is the speed of light. Simulations of moving shocks require large computational domains. This can easily make the computational burden unfeasible, especially when running simulations with high spatial and temporal resolution, in three dimensions, and/or when employing significant numbers of numerical particles, usually required to better capture the ion microphysics. 
}

Collisionless plasma shocks in kinetic simulations are typically initialized with the ``piston method'' \citep{winske1988}. Assuming the ``solar wind'' plasma flows in a simulation domain from left to right with a super-Alfv\'enic Mach number, $M_A=V_{SW}/V_A>1$, where $V_{SW}$ and $V_A$ are the solar wind and Alfv\'en speeds, a simulation shock is produced by continuously injecting the plasma from the left simulation boundary and reflecting it off the right boundary, so that the backstreaming flow continues to interact with the incoming plasma. This (``downstream frame'') setup produces a shock, propagating back towards the plasma injection boundary, and a thermalized plasma at rest downstream of the shock. Alternatively, a moving shock may be initialized in the ``upstream frame'' by continuously injecting a super-Alfv\'enic flow into the plasma that remains at rest in the upstream (simulation) frame of reference. In either case, one has to follow a numerical shock that propagates through the computational domain.
\textcolor{myred}{The domain, therefore, needs to be large enough in the direction of shock propagation, to enable long simulation times required for convergence of physical results of interest (if such convergence can be achieved at all).}

Furthermore, as the shock moves across the simulation box, the distance between its front and the injection boundary inevitably shortens with time. The continuous reduction of the length of the foreshock region limits the time available for plasma instabilities to generate upstream turbulence \citep{bennett1995, simon2026}, which also negatively impacts the convergence of simulation results. Since reflecting ions gain energy when they scatter back and forth at the shock front, before leaving the simulation through the upstream boundary, the shorter the foreshock region, the less energy a particle can gain before it is lost from the simulation. As a result, as the shock approaches the upstream boundary, an increasing fraction of energetic particles is removed from the simulation and their total energy density decreases with time, which is opposite to what one would expect to happen in the presence of stochastic acceleration \citep{bennett1995}.

\textcolor{myred}{The motion of a plasma shock through the simulation domain remains intrinsically transitory, unless simulation time is long enough. This complicates studying the physics of shocks, including the efficiency of ion acceleration. This is true especially for quasi-parallel and oblique shocks, where shock reformation is known to be inherently unsteady \citep{bennett1995}.} Increasing the length of the simulation box inevitably makes investigation of shock behavior and ion acceleration very expensive, especially in three dimensions. In hybrid simulations, to improve convergence of energetic ion distributions, an upstream free escape boundary (FEB) may be used, usually kept at a fixed distance upstream from the moving shock front \citep{giacalone1997}. In this case, ions, moving upstream and crossing this boundary, are removed from the simulation. The simulation must stop when the FEB reaches the simulation upstream boundary; the only way to increase the run time then is to increase the simulation domain length. 

\textcolor{myred}{The FEB length in self-consistent hybrid simulations of high-energy ions (with no external upstream perturbations) must be large enough to resolve their mean-free paths, which are usually much greater than their gyro-radii. This scale determines the minimum physical upstream length required in such simulations. One typically needs to run the simulation of a moving shock for a very long time just to establish convergence of shock upstream properties. Alternatively, to speed up shock formation and convergence of ion spectra, external magnetic perturbations may be introduced \citep{Giacalone1992}. The transverse domain dimension must also be chosen to be sufficiently large in order to resolve large particle gyro-radii and allow relevant ion microphysics to develop and evolve on appropriate physical scales \citep{simon2026}.  
}

While the shock is moving across the simulation domain, large parts of its upstream and downstream portions remain relatively ``inactive'' due to smaller particle densities and slower field changes, while interactions in the shock transition region must be accounted for with finer temporal resolution and with more particles. For non-adaptive domain decompositions, this may significantly degrade parallel performance of hybrid simulations. 

The above numerical issues make hybrid (and fully kinetic) simulations of moving shocks computationally expensive. In addition, simulation results may not always be conclusive. For instance, to investigate the efficiency of shock-driven ion acceleration in hybrid simulations, one needs to explore a large space of parameters, including the shock Alfv\'enic Mach number, $M_A$, magnetic inclination angle, $\theta_B$ (obliquity), and upstream values of plasma density, temperature, and background magnetic field. The success of these parametric studies relies heavily on performing many high-fidelity simulations, which also may require small time steps to maintain numerical accuracy \citep{caprioli2014}. This necessitates exploration of alternative, more efficient methods for kinetic simulations of collisionless plasma shocks.

\subsection{Shock-Frame Simulations}

\textcolor{myred}{Strict requirements for spatial and temporal accuracy of kinetic plasma simulations impose severe computational constraints on simulations of moving shocks. Currently, multi-dimensional hybrid simulations are highly restrictive, while full-electromagnetic PIC modeling remains computationally prohibitive. Naturally, this predicament necessitates development of computational techniques for simulations of shocks in their co-moving (shock) frame. In the past, multiple attempts have been made to achieve this goal. Contrary to magnetohydrodynamics (MHD), however, initialization of quasi-stationary kinetic shocks with the Rankine-Hugoniot conditions does not appear to provide robust means for establishing stable shocks over long ion timescales, as became evident from earlier one-dimensional full-electromagnetic \citep{umeda2011} and hybrid \citep{leroy1981} simulations of quasi-perpendicular shocks, run only for relatively short periods of time compared to the ion gyroperiod. 
}

Hypothetically, a self-consistent shock-frame simulation may be achieved by modifying the downstream boundary to allow the incoming plasma to ``freely flow'' through the simulation box. For instance, this technique was used in one-dimensional hybrid simulations, where plasma ions were passed through or reflected off the right boundary in a probabilistic manner \citep{bennett1995}. It was found, however, that the efficiency of this method in establishing a stationary shock front greatly depends on guesswork regarding the expected density compression ratio. Nevertheless, once the stationary front position was obtained in the shock frame, a 1D hybrid simulation could run for a long time ($\sim 1000\,\Omega_{ci}^{-1}$), using a small simulation box and a relatively small number of particles \citep{bennett1995}. 

\textcolor{myred}{Recently, a new ``faux-shock'' frame simulation technique has been proposed to enable more efficient multi-dimensional hybrid simulations of shock precursor driven CRs \citep{simon2026}. Similar to the technique used by \citet{bennett1995}, the ``faux-shock'' method introduces a semi-transparent outflow boundary, so that ions are probabilistically allowed to pass through or forced to be reflected off this boundary, based on the relation of their speeds to the downstream plasma speed. The ``faux-shock'' setup, however, does not self-consistently describe the shock evolution in the downstream region and excludes the complex plasma transition at the shock front. In what follows, we describe how to resolve these critical issues and achieve self-consistent hybrid simulations of shocks and ion acceleration in the shock frame. Our method can equally be used to investigate ion acceleration in both the upstream and downstream regions. This effectively corresponds to studies focusing on astrophysical CRs and heliospheric seed ions, respectively.
}

\section{An Adaptive Frame-of-Reference Algorithm (AFORA) for Shock Simulations} \label{sec:AFORA}

\textcolor{myred}{In this Section, we describe a novel computational method for enabling long self-consistent hybrid simulations of non-relativistic shocks in computational domains with relatively small lengths in the direction of plasma flow. We demonstrate this method for shocks with different obliquities, $\theta_B$. This is achieved by generating a shock with the ``piston method'' and then transitioning the simulation to the co-moving shock frame, where the reflecting wall is replaced with a self-adaptive ``free'' downstream boundary, different from that proposed by \citet{bennett1995}.
}

We begin a simulation by reflecting the incoming plasma flow off the stationary right simulation boundary. As a ``piston-generated'' shock propagates to the left, we determine its speed, using two methods for tracking the position of its front, $x_\mathrm{sh}$. The first, ``current-based'' algorithm assumes the front position to coincide with the cell position, $x_k$, corresponding to the maximum absolute value of the total current, $\bf J = \nabla \times {\bf B}$:
\begin{equation}
  x_\mathrm{sh} = \{ x_k \rightarrow |{\bf J}|_{max} \} 
  \label{eq:xsh1}  
\end{equation}
The second, ``density-based'' algorithm approximates the upstream and downstream plasma volumes with rectangular boxes and determines $x_\mathrm{sh}$ by estimating the position of a center of mass, $x_\mathrm{cm}$:
\begin{equation}
x_\mathrm{sh}=2x_\mathrm{cm}-x_\mathrm{rb}, 
x_\mathrm{cm}= \frac {\sum_k (n_k-n_u)x_k}{\sum_k(n_k-n_u)}
\label{eq:xsh2}
\end{equation}

In Eqs.~(\ref{eq:xsh1}-\ref{eq:xsh2}), $x_\mathrm{rb}$ is the right (downstream) simulation boundary position, $n_k$ is the ion number density at cell position, $x_k$, and $n_u$ is the upstream number density. 
\textcolor{myred}{The current-based front tracking algorithm has been found to be more accurate in the initial (``piston'') stage of shock evolution, while the density-based tracking algorithm produces smoother results for calculating adaptive front displacements in the shock frame. In other words, we use Eq.~(\ref{eq:xsh1}) for tracking $x_\mathrm{sh}$ in the downstream frame and Eq.~(\ref{eq:xsh2}) thereafter, after applying the Lorentz transformation of the simulation system into the shock frame (see below).
}

\textcolor{myred}{Since the shock front may be turbulent, we break the computational domain into a set of subdomains parallel to the x-axis. At each given time, individual shock front positions, found in these subdomains, are averaged to deduce the final front position, $x_\mathrm{sh}$. Shock position queries are performed at equal time intervals, $\Delta t_\mathrm{sh}$. In the downstream frame, the shock front tracking algorithm is used to estimate the current shock speed, time-averaged for a final value. In the shock frame, the front tracking routine computes shock front displacements used for adaptive control of the ``semi-transparent'' downstream  boundary, as explained below.
}

\textcolor{myred}{After the shock front has moved in the downstream frame to approximately the middle of the computational domain and its velocity in the x-direction, $V_\mathrm{sh,0}<0$, has been computed, the simulation system is converted to the shock frame using a slightly modified shock velocity, $V_\mathrm{sh}=K(\theta_B)V_\mathrm{sh,0}<0$,
where $K(\theta_B)\geq 1$ is an empirical computational factor, expressed as a function of shock obliquity angle, $\theta_B$. For smaller values of $\theta_B$, choosing $K(\theta_B)=1$ results in stopping the shock past the designated position, closer to the plasma injection boundary. Choosing $K(\theta_B)>1$ for these angles (see below) makes the shock front stop closer to the middle of the domain. This finding correlates with smaller shock speeds measured for smaller angles, $\theta_B$ \citep{preisser2020}.
}

\textcolor{myred}{At this moment in time, the modified shock velocity, $V_\mathrm{sh}$, is subtracted from all particle velocities. The transformation of ion velocities, together with the accompanying transformation of the electron fluid velocity (Ampere's law), naturally results in the corresponding Lorentz transformation of the electric field, $\bf E$, found from the electron equation of motion (Ohm's law) in the hybrid model. Accordingly, from this moment on, ions are injected from the left simulation boundary with a modified injection velocity, $V_u'=V_u-V_\mathrm{sh}$, as well. Currently, the magnetic field, $\bf B$, is not Lorentz-transformed, since magnetic field corrections are not significant for non-relativistic shocks.
}

The rest of the algorithm executes the following intuitive idea. 
\textcolor{myred}{For quasi-stationary upstream conditions (including external upstream turbulence), the shock front position must also remain quasi-stationary in the shock frame. 
}
To enforce this behavior, we apply an adaptive downstream boundary condition that converts the previously reflective boundary to a ``free'' one. The logic of this adaptivity can be understood as follows. Suppose this boundary is left purely reflective after the Lorentz transformation. Then, the shock front will continue moving towards the injection boundary, being pushed by new reflected particles that reach the front. On the other hand, if the right boundary is made purely absorbing, the shock front will start moving towards the outflow boundary due to the drop of plasma pressure in the downstream region.

Therefore, once the simulation restarts in the shock frame, we make the right simulation boundary ``semi-transparent'' for particles. Specifically, an outgoing particle experiences a specular reflection at the outflow boundary if the plasma density at its cell is found to be below a certain threshold density value, $n_d$; otherwise, the particle is absorbed. Assuming zero post-shock plasma velocity, $V_d=0$, in the downstream frame, we estimate the corresponding value of $n_d$ through the mass continuity equation:
\begin{equation}
n_u (V_u-V_\mathrm{sh})= n_d ((V_d=0)-V_\mathrm{sh}) \implies
n_d= n_u(M_A+1),
\label{eq:cont}
\end{equation}
where $M_A=V_u/\left | V_\mathrm{sh} \right |$ is the Alfv\'enic Mach number for the upstream flow in the downstream frame. 

\textcolor{myred}{Following initialization, the value of $n_d$ is adaptively modified by the shock query algorithm at time intervals, $\Delta t_\mathrm{sh}$, to ensure the shock front remains quasi-stationary. This is done by tracking the shock front displacement, $\Delta x_\mathrm{sh}$, and replacing the previous value of the threshold density, $n_d^{old}$, with a new value, $n_d^{new}$, as follows:
}
\begin{equation}
    n_d^{new} = n_d^{old}e^{\alpha \Delta x_\mathrm{sh}}
    \label{eq:exp}
\end{equation}

\textcolor{myred}{In Eq.~(\ref{eq:exp}) $\alpha=5\times 10^{-3}$ is selected so that $\alpha|\Delta x_\mathrm{sh}| \ll 1$.}
It follows from this equation that $n_d$ increases when the shock front shifts towards the outflow boundary ($\Delta x_\mathrm{sh}>0$). This forces more particles to be reflected, which pushes the front back towards the injection boundary. In the opposite scenario ($\Delta x_\mathrm{sh}<0$), $n_d$ is reduced, enabling more particles to pass through the outflow boundary, which makes the shock front shift in the opposite direction, towards the outflow boundary.

\textcolor{myred}{The nonlocal control of the outflow boundary distinguishes our approach from those taken by \citet{bennett1995} and \citet{simon2026}. The criterion for reflection (absorption) of particles in our study is not based on a probabilistic relation of their speeds to a fixed downstream flow speed, which \citet{bennett1995} had to find by trial and error. Instead, we formulate this criterion in terms of the threshold density, $n_d$, initially computed from the mass continuity equation and adaptively modified thereafter as explained above. We show below (Section~\ref{sec:disc}) that our algorithm produces satisfactory results even when the shock front becomes rippled and turbulent. 
}

\textcolor{myred}{Eq.~(\ref{eq:cont}) does not take into account shock-reflected particles that arise in supercritical shocks and contribute to overall mass balance. 
Therefore, to keep the shock front quasi-stationary, one inevitably needs to control its position adaptively, as presented above, by accounting for the changing fraction of backstreaming particles that may still be evolving during the transition between the downstream and shock frames. When a quasi-steady state is finally achieved in the shock frame, the number of particles in our simulations, $N_p$, remains roughly the same, fluctuating in time in direct correspondence with the fluctuating shock position, $x_\mathrm{sh}$ (see Figure~\ref{fig:diagnostics}).
}

\section{Simulation Setup}
\label{sec:HYPERS}

\textcolor{myred}{The AFORA approach has been implemented in a multiscale hybrid code, HYPERS (Hybrid Parallel Event-Resolving Simulator) \citep{Omelchenko2012a}. We refer to this implementation as the HYPERS-Shock model. HYPERS solves a standard set of hybrid model equations, discretized on a logically uniform mesh that may be uniform or stretched in physical space \citep{Omelchenko2021a}. In this study, we use uniform meshes only. We emphasize, however, that the accuracy and efficiency of quasi-stationary shock simulations in the shock frame can further be improved by refining a stretched mesh around the shock front where critical mechanisms for ion acceleration take place. Similar mesh refinement would be difficult to implement for ``piston-generated'' shocks because of their non-stationary nature.  
}

Compared to conventional (time-stepped) hybrid codes, asynchronous updates of particles and electromagnetic fields in HYPERS are enabled by a multiscale simulation technology, EMAPS (Event-driven Multi-Agent Planning System) \citep{Omelchenko2023spr}. EMAPS functions as a ``simulation time operating system'', converting a physical model into a multiscale computational system, driven by intelligent, asynchronously interacting, parallel agents (cells). EMAPS adaptively assigns dynamic time increments to fields and particles, effectively ensuring physically small but finite changes to their properties per each update. This event-driven time integration has been shown to achieve superior numerical accuracy in realistic (nonlinear, stochastic, coupled) systems, compared to equivalent time stepping methods \citep{Omelchenko2006, Omelchenko2023spr}. The latter are known to suffer from numerically diffusive and unstable calculations, when too small and too large timesteps are used, respectively. Earlier applications of the event-driven integration to 1D hybrid simulations of plasma shocks and discontinuities demonstrated the self-adaptive methodology to be more accurate and faster compared to equivalent explicit and implicit time-stepping techniques \citep{Omelchenko2006a}.

In this paper, we demonstrate the flexibility and accuracy of AFORA by performing a series of 2D ($xy$) simulation runs, conducted with identical physical inputs, except for different shock obliquities (angles between the ambient magnetic field, $\bf {B}_0$, and the solar wind plasma flow direction), $\theta_B=0^\circ, 20^\circ, 40^\circ, 60^\circ, 80^\circ$, for which we use $K(\theta_B)=1.3, 1.2, 1.1, 1.0, 1.0$, respectively (see Section~\ref{sec:AFORA}).

To initialize simulations, the following dimensionless parameters have been used for the upstream plasma: 1) the ratio of the ion (proton) plasma frequency ($\omega_{pi}$) to the cyclotron frequency ($\Omega_{ci}$), $\omega_{pi}/\Omega_{ci}=c/V_A=6,000$, 2) the Alfv\'enic Mach number, $M_A=V_u/V_A=3$ ($V_u$ is the upstream plasma flow speed in the downstream frame, where $V_d=0$), and 3) the ion and electron thermal to magnetic field energy density ratios, $\beta_i=0.383$, $\beta_e=0.5$, respectively. These values are within the range of parameters of commonly observed interplanetary shocks \citep{gargate2012,preisser2020}.

All spatial scales are normalized to the upstream ion inertial length, $d_i=c/\omega_{pi}$, and simulation time is measured in $1/\Omega_{ci}$. Plasma densities and velocities are normalized to the upstream plasma density and Alfv\'en speed, $V_A$, respectively. Ion (proton) temperatures are normalized by $m_pV_A^2$ ($m_p$ is the proton mass), and magnetic field magnitudes are normalized to the upstream magnetic field strength, $|\bf B_0|$. Energies of shock-accelerated protons (see Section~\ref{sec:disc}) are computed in the solar wind frame and normalized to the corresponding proton injection energy, $E_0=m_p(V_u-V_\mathrm{sh})^2/2$. All simulation runs have been performed on a uniform mesh, $N_x\times N_y=400\times 400$ cells, with equal cell sizes, $\Delta x=\Delta y=0.5 d_i$, initialized with 25 ion macro-particles per cell. The computational domain is periodic in the y-direction.

To filter out the short-scale (of the order of cell size) magnetic field perturbations, we use a nonuniform resistivity, $\eta$ \citep{Omelchenko2021a}, chosen to be small enough to guarantee large numerical magnetic Reynolds numbers at all cells:
\begin{equation}
    Rm= \frac{\Delta t_d}{\Delta t_A}=
    \frac{2\pi\Delta x}{\eta c}\frac{V_A}{c}\geq12,
\end{equation}
\textcolor{myred}{where $\Delta t_d$ and $\Delta t_A$ are the characteristic magnetic diffusion and Alfv\'enic transit times across the computational cell, respectively. In hybrid simulations, the magnetic Reynolds number, $Rm$, must be ensured to be large enough for numerical wave effects to dominate numerical resistive effects.
}

\textcolor{myred}{Simulation ions are continuously injected from the upstream (left) boundary, randomly generated using a drift-Maxwellian velocity distribution with the bulk drift velocity, $V_u$, directed along the positive x-direction. 
As explained above (Section~\ref{sec:AFORA}), the downstream (right) boundary initially acts as a perfectly reflecting wall. The interaction between the reflected and incoming flows thermalizes the downstream plasma ($V_d\simeq 0$) and produces an initial ``piston-generated'' shock that propagates in the negative x-direction. Once the shock front reaches the middle of the computational domain, the simulation system is transformed into the shock frame. This transformation takes place very early in time, $\Omega_{ci}t\sim 100$. From this moment on, the right boundary acts as a free outflow boundary. 
}

\begin{figure}[!ht]
\centering
\includegraphics[width=\textwidth]{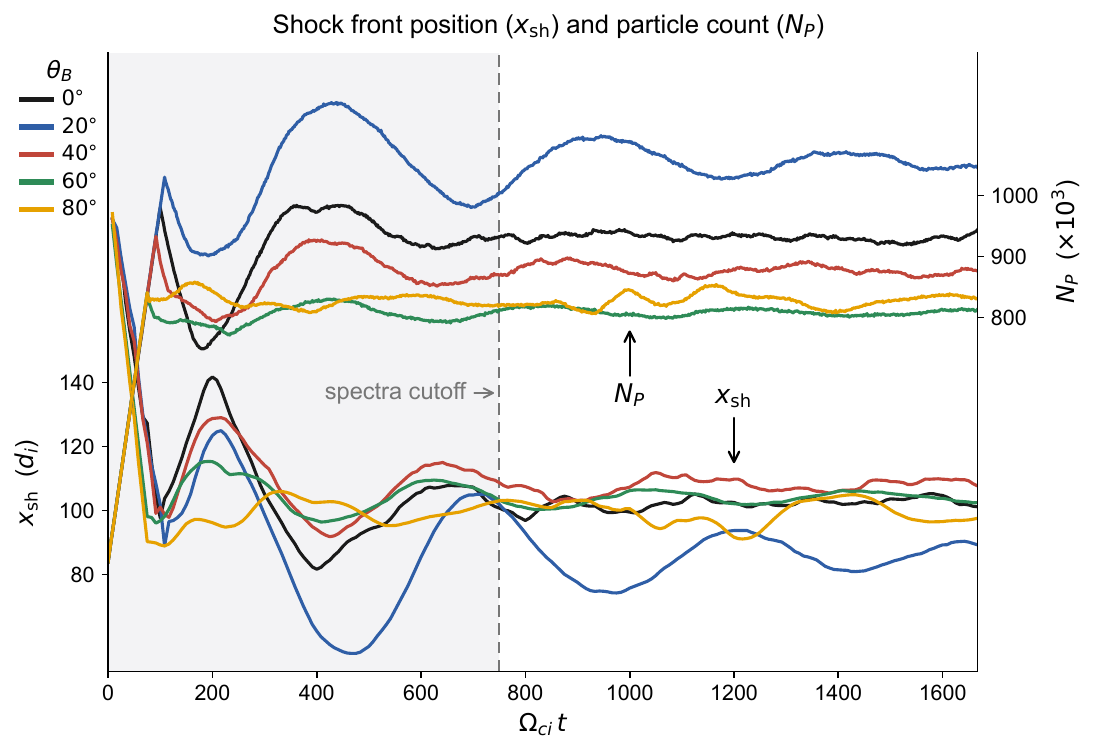}
\caption{Stationarity of AFORA shocks, demonstrated with two run-level diagnostics for all five inclination angles ($\theta_B=0^\circ$--$80^\circ$; indicated by a legend at the upper left corner) on the shared time axis, $\Omega_{ci}t$. 
Lower band (left vertical axis): the shock-front position, $x_\mathrm{sh}$, dynamically tracked during the simulation.  
Upper band (right vertical axis): the total number of randomly selected simulation particles, $N_p$ ($\times10^3$). Each value of $\theta_B$ uses a single color shared by both groups of curves. The two groups are offset into separate bands purely for legibility, and a black arrow labels each band.
}
\label{fig:diagnostics}
\end{figure}

For all values of $\theta_B$ used, Figure~\ref{fig:diagnostics} demonstrates two run-time diagnostics used to verify the long-time stationarity of the simulation system: the tracked shock-front position, $x_\mathrm{sh}$, and the number of particles (at all times selected from the full set with the same probability), $N_p$. During the initial transient period (shaded region, $\Omega_{ci}t<750$), the front position, $x_\mathrm{sh}$, swings as the shock relaxes into the shock frame following the Lorentz transformation; past this time, $x_\mathrm{sh}$ settles onto an asymptotic value for every angle, confirming that the adaptive outflow boundary keeps the shock front stationary in the simulation box, rather than letting it drift toward either simulation boundary. Over the same interval of time, the number of simulation particles, $N_p$, nearly flattens, showing that AFORA balances the injected upstream flux so that the number of particles in the system  neither gets progressively depleted nor accumulated over the long run time, $\sim 1600\,\Omega_{ci}^{-1}$.
 
The oscillation amplitudes of these diagnostics may be reduced in the future by tuning computational parameters of the HYPERS-Shock model more optimally. Note that the dashed line in Figure~\ref{fig:diagnostics} marks a ``cutoff time'', $\Omega_{ci}t=750$, following which spectra of energetic particles are computed (see more in Sec.~\ref{sec:disc}), with the shaded region ($\Omega_{ci}t<750$) representing the front-settling time period, discarded when computing spectral averages in Figure~\ref{fig:spectra} and time averages of simulation quantities in Table~\ref{tab:runs}. 

\section{Discussion of Simulation Results}
\label{sec:disc}

\textcolor{myred}{In this Section, we discuss three important aspects of HYPERS-Shock simulations: (i) quasi-stationarity of shock behavior in the shock frame, (ii) physical consistency of our numerical shocks, compared to space observations, and (iii) quasi-steady spectra of downstream energetic ions, to be used for coupling the HYPERS-Shock model with Fokker-Planck models that predict the intensity of SEP events at Earth's orbit \citep{sokolov2004new, liu2025physics}. Our findings are further illustrated with animations of plasma shocks with different obliquities, $\theta_B$ \citep{Shockmovies2026}. The purpose of these animations is to demonstrate the successful implementation of AFORA in HYPERS and help the reader fully grasp the transition of a hybrid simulation from its initial ``piston'' stage (downstream frame) to the shock frame, where plasma dynamics remains quasi-stationary over long simulation times. The shock-frame simulation offers two clear advantages over the standard ``piston'' method: 1) quasi-stationary, converged plasma behavior, and 2) reduction of the length of a simulation domain.  
}

\subsection{Quasi-Stationarity of Shock Profiles in the Shock Frame}

\begin{figure}[!ht]
\centering
\includegraphics[width=\textwidth]{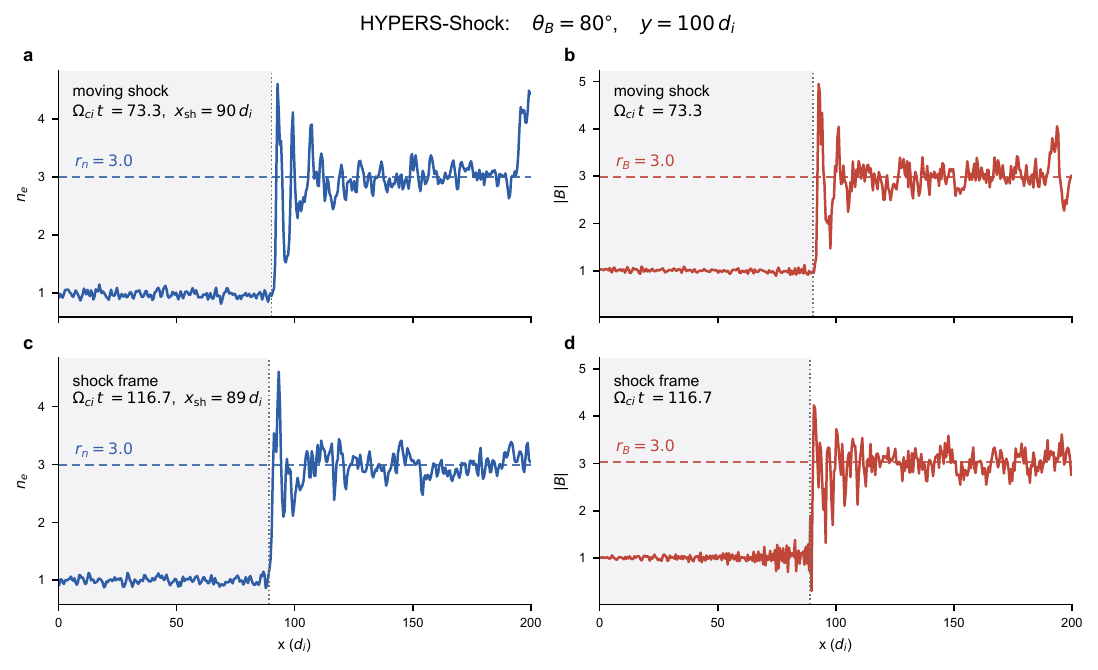}
\caption{
\textcolor{myred}{Transformation of the quasi-perpendicular shock, $\theta_B=80^\circ$, from the downstream frame to the shock frame, illustrated with 1D cuts along $y=100\,d_i$. The top row shows the last ``piston-generated'' moving shock at $\Omega_{ci}t=73.3$: \textbf{(a)} electron density, $n_e$, and \textbf{(b)} magnetic-field magnitude, $|\mathbf{B}|$. The bottom row shows the corresponding shock-frame profiles at $\Omega_{ci}t=116.7$: \textbf{(c)} $n_e$ and \textbf{(d)} $|\mathbf{B}|$. The dotted vertical lines mark the shock front, and the dashed horizontal lines indicate the downstream compression levels, $r_n=r_B=3.0$. Soon after the frame transformation, the magnetic-field profile develops enhanced fluctuations immediately upstream of the front.}
}
\label{fig:frame_transform_n80}
\end{figure}

\begin{figure}[!ht]
\centering
\includegraphics[width=\textwidth]{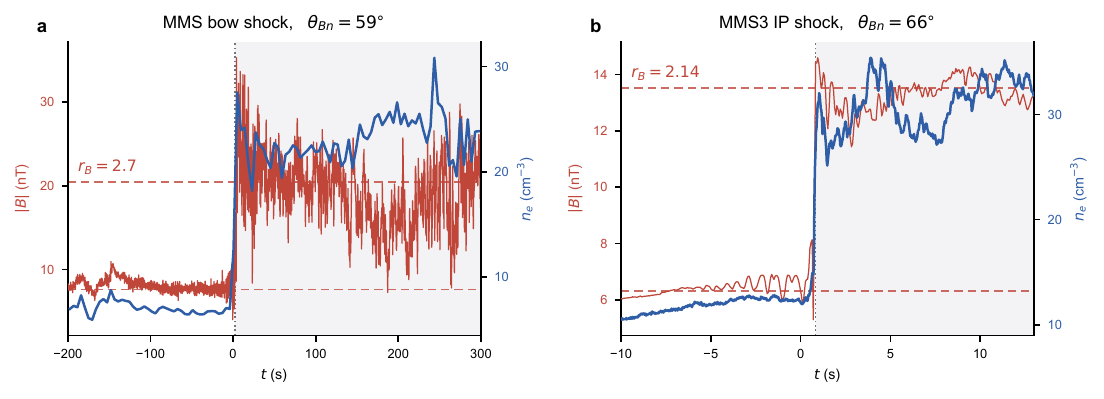}
\caption{
\textcolor{myred}{Two shock crossings spanning the oblique-to-quasi-perpendicular range, selected from the SHARP Shock Database: \textbf{(a)} Earth's bow-shock crossing observed by MMS on 2017 October 12 at approximately 15:22:29 UT, and \textbf{(b)} an interplanetary-shock crossing observed by MMS3 on 2018 January 8 at approximately 06:41:10 UT. In both panels the time coordinate is referenced to the crossing, the dotted vertical line marks the identified shock transition, and the shaded region behind it is downstream. The red curve (left axis) shows the magnetic-field magnitude, $|B|$, and the blue curve (right axis) shows the electron number density, $n_e$. The red dashed horizontal lines indicate the mean upstream and downstream magnetic-field magnitudes, $B_u$ and $B_d$, respectively, whose ratio, $r_B$, is annotated on the downstream level and is directly comparable to the $r_B$ column of Table~\ref{tab:runs}. In panel (a), the shock-normal angle is $\theta_{Bn}\simeq59^\circ$, corresponding to an oblique shock, and the present analysis gives $M_A\simeq4.95$. In panel (b), the shock-normal angle obtained from mixed coplanarity is $\theta_{Bn}\simeq66^\circ$, corresponding to a quasi-perpendicular shock, and the present analysis gives $M_A\simeq1.95$.} 
}
\label{fig:observations}
\end{figure}

\begin{figure}[!ht]
\centering
\includegraphics[width=\textwidth]{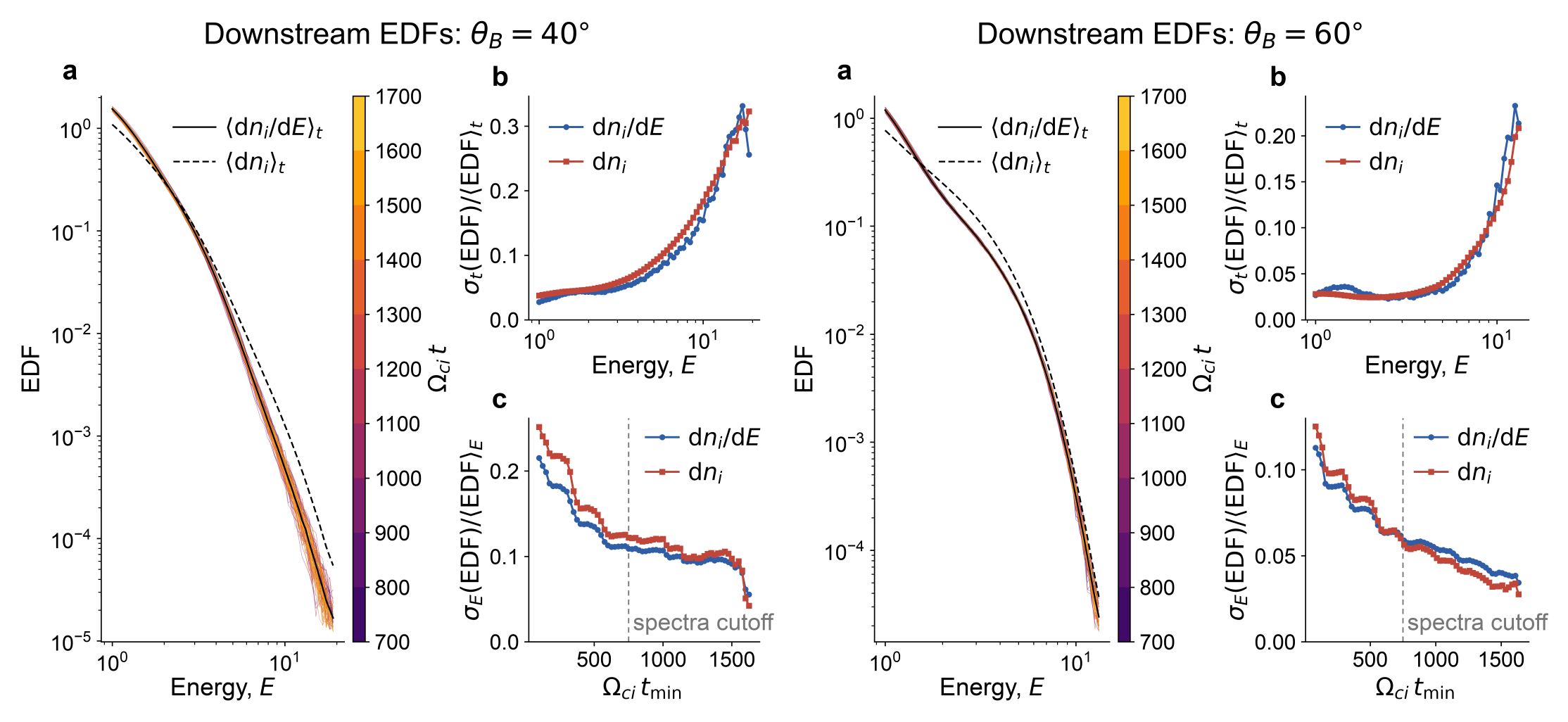}
\caption{Stationarity of downstream ion energy distributions, demonstrated for two representative angles, $\theta_B=40^\circ$ (left block) and $\theta_B=60^\circ$ (right block). Each block has three panels: \textbf{(a)} energy distribution functions (EDFs), recorded in the quasi-steady window, $\Omega_{ci}t\ge750$, represented by one thin line per simulation shot, colored by shot time, and overlaid with the time-averaged differential spectrum $\langle\mathrm{d}n_i/\mathrm{d}E\rangle_t$ (solid black) and the time-averaged cumulative spectrum $\langle\mathrm{d}n_i\rangle_t$ (dashed black), with per-shot curves collapsing onto the mean curve at later times; \textbf{(b)} shot-to-shot relative scatter error, $\sigma_t(\mathrm{EDF})/\langle\mathrm{EDF}\rangle_t$ for the differential ($\mathrm{d}n_i/\mathrm{d}E$, blue) and cumulative ($\mathrm{d}n_i$, red) spectra: the error stays small ($\lesssim5\%$) and flat across the well-sampled energy range and grows only in the sparse high-energy tail; \textbf{(c)} the energy-averaged scatter error, $\sigma_E(\mathrm{EDF})/\langle\mathrm{EDF}\rangle_E$, as a function of the start time of the averaging window, $\Omega_{ci}t_{\min}$: the error decreases and then plateaus, showing that the spectra converge once the early, still-moving-front shots are dropped. The dashed line marks the fixed cutoff time, $\Omega_{ci}t=750$ (``spectra cutoff'') used throughout, including for defining the time window over which the ion acceleration efficiency entries in Table~\ref{tab:runs} are averaged. 
}
\label{fig:spectra}
\end{figure}

\textcolor{myred}{Simulation results, obtained for a range of obliquity angles, $\theta_{B}=0^\circ$--$80^\circ$, indicate that the AFORA method enables studying quasi-stationary shocks over simulation periods, $t \sim 1600 \Omega_{ci}^{-1}$, long enough to characterize important turbulent processes and concomitant ion acceleration at shocks. During most of the simulation period, shock structures and distributions of energetic ions (see below) remain quasi-steady in the shock frame (Figures~\ref{fig:diagnostics}--\ref{fig:spectra}), which proves the numerical robustness of our method.
}

\begin{figure}[p]
\centering
\includegraphics[width=\textwidth]{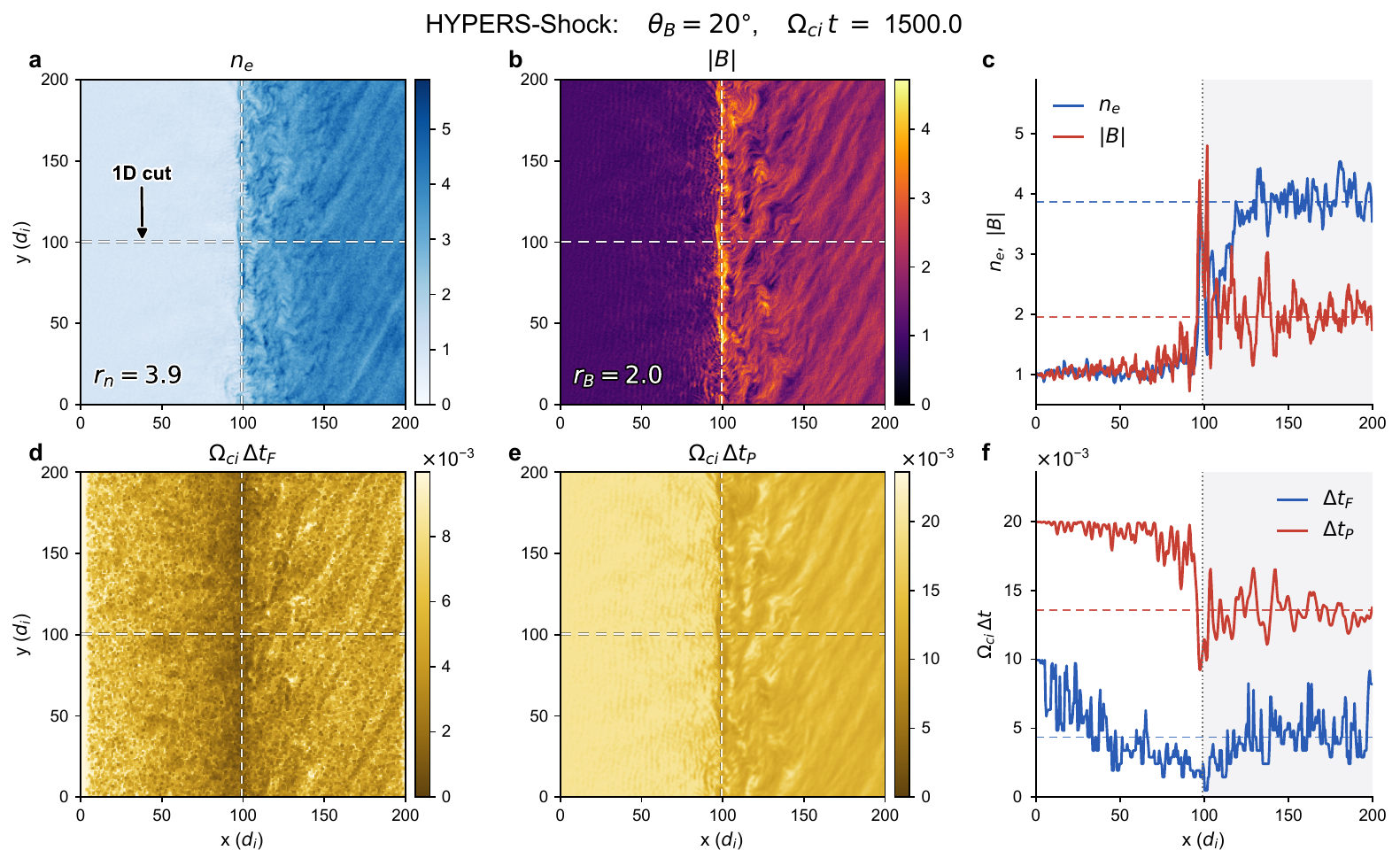}
\caption{The structure of the 2D quasi-parallel shock for $\theta_B=20^\circ$ at $\Omega_{ci}t=1500$: \textbf{(a)} electron density $n_e$ and \textbf{(b)} magnetic-field magnitude $|\mathbf{B}|$, with instantaneous compression ratios, $r_n$ and $r_B$; \textbf{(c)} 1D cuts of $n_e$ (blue) and $|\mathbf{B}|$ (red) along the horizontal dashed line at $y=100\,d_i$ (see the arrow in panel~a), with the downstream averages dashed and the downstream region shaded. \textbf{(d,e)} self-adaptive time increments for fields $\Omega_{ci}\Delta t_F$ and particles $\Omega_{ci}\Delta t_P$, with \textbf{(f)} their 1D cuts. The dashed vertical line marks the dynamically tracked shock front. The quasi-parallel shock front is broad and rippled, with extended upstream wave activity from reflected ions. The field and particle time increments contract at the front and in the downstream region.
}
\label{fig:quad20}
\end{figure}

\begin{figure}[p]
\centering
\includegraphics[width=\textwidth]{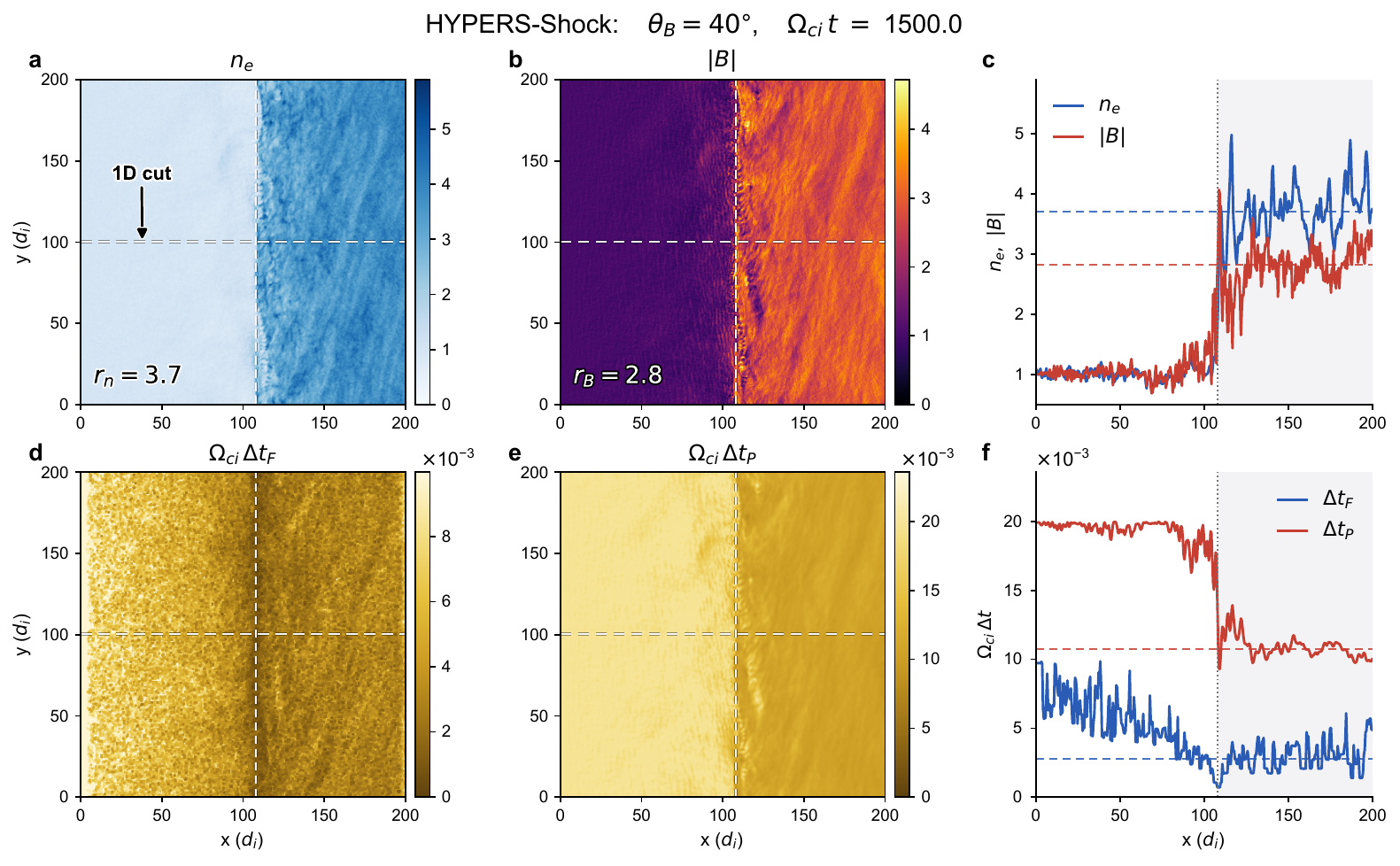}
\caption{Same as Figure~\ref{fig:quad20}, except for the oblique shock angle, $\theta_B=40^\circ$. The front transition is sharper than at $\theta_B=20^\circ$ and the magnetic field compresses more strongly ($r_B\simeq2.8$) while the density compression stays near $r_n\simeq3.7$. The adaptive time increments again shrink at the front and through the downstream.}
\label{fig:quad40}
\end{figure}

\begin{figure}[p]
\centering
\includegraphics[width=\textwidth]{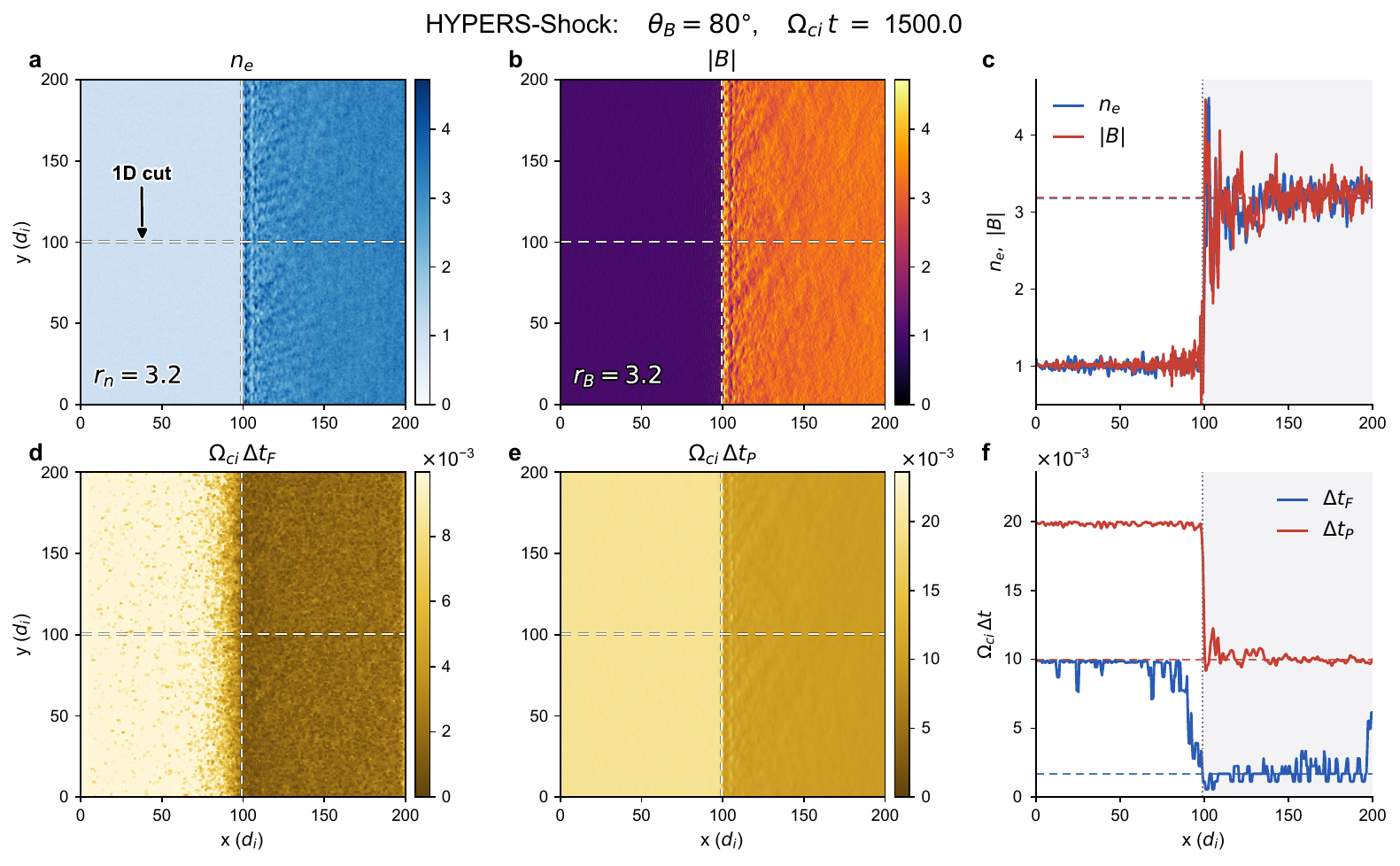}
\caption{Same as Figure~\ref{fig:quad20}, except for the quasi-perpendicular shock angle, $\theta_B=80^\circ$. The shock front is a thin, nearly one-dimensional layer, across which $n_e$ and $|\mathbf{B}|$ jump together ($r_n\simeq r_B\simeq3.2$), with little upstream wave activity. The adaptive field increments drop by an order of magnitude across the sharp front transition.}
\label{fig:quad80}
\end{figure}

\begin{figure}[p]
\centering
\includegraphics[width=\textwidth]{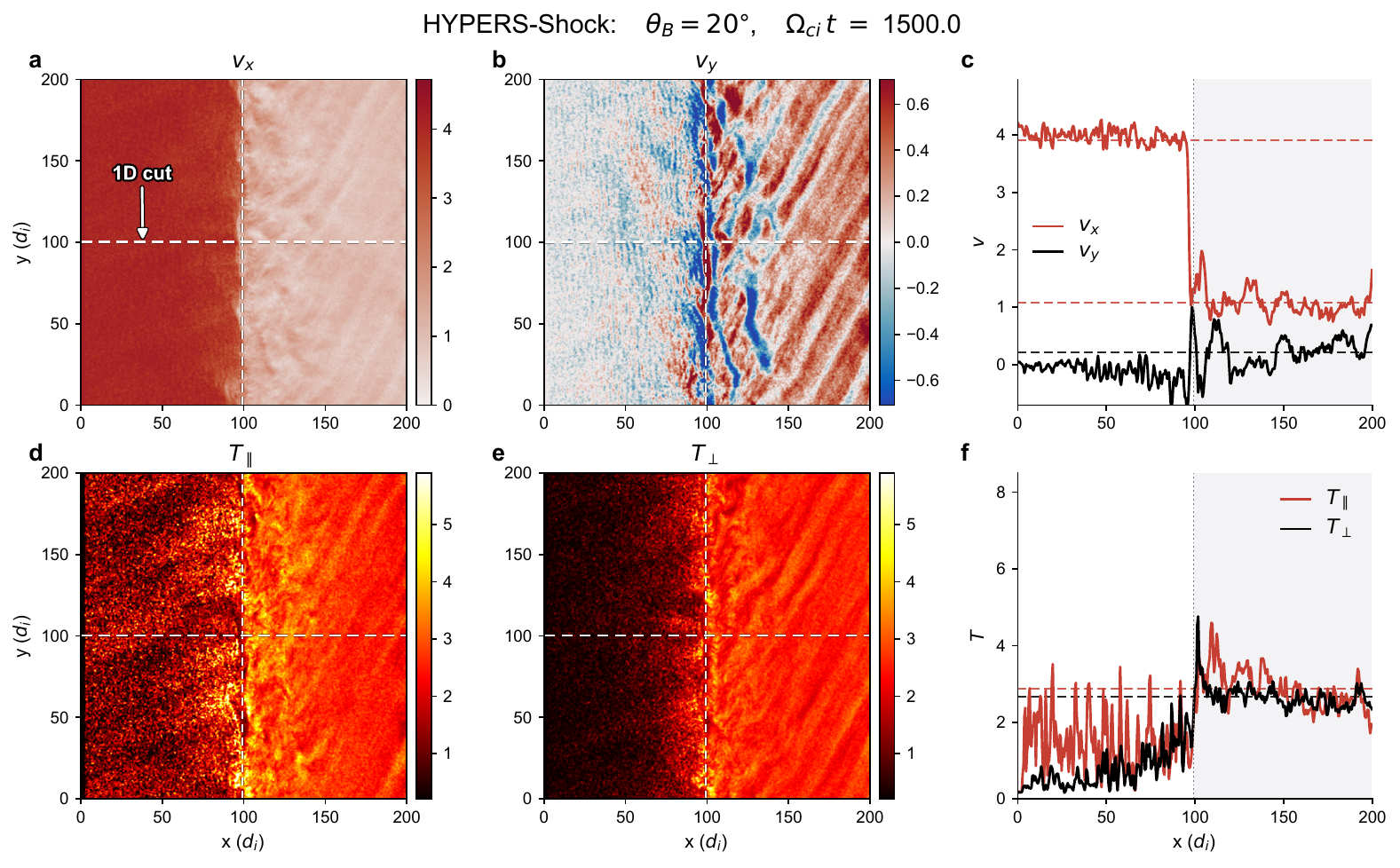}
\caption{Ion flow velocities and temperatures of the quasi-parallel shock, $\theta_B=20^\circ$ at $\Omega_{ci}t=1500$, corresponding to Figure~\ref{fig:quad20}: \textbf{(a)} velocity component, $v_x$, \textbf{(b)} velocity component, $v_y$, \textbf{(c)} their 1D cuts along $y=100\,d_i$; \textbf{(d)} parallel temperature, $T_\parallel$, \textbf{(e)} perpendicular temperature, $T_\perp$ (shared scale), \textbf{(f)} their 1D cuts. 
The dashed vertical line marks the shock front and the dashed horizontal lines indicate the downstream averages, with a second dashed line in the $v_x$ cut (panel c) marking the upstream inflow speed. The upstream inflow velocity component $v_x\simeq4\,V_A$ decelerates across the front. The structured downstream flow with velocity component $v_y$ and strong, field-aligned upstream ion beams are characteristic of the quasi-parallel shock geometry.}
\label{fig:mom20}
\end{figure}

\begin{figure}[p]
\centering
\includegraphics[width=\textwidth]{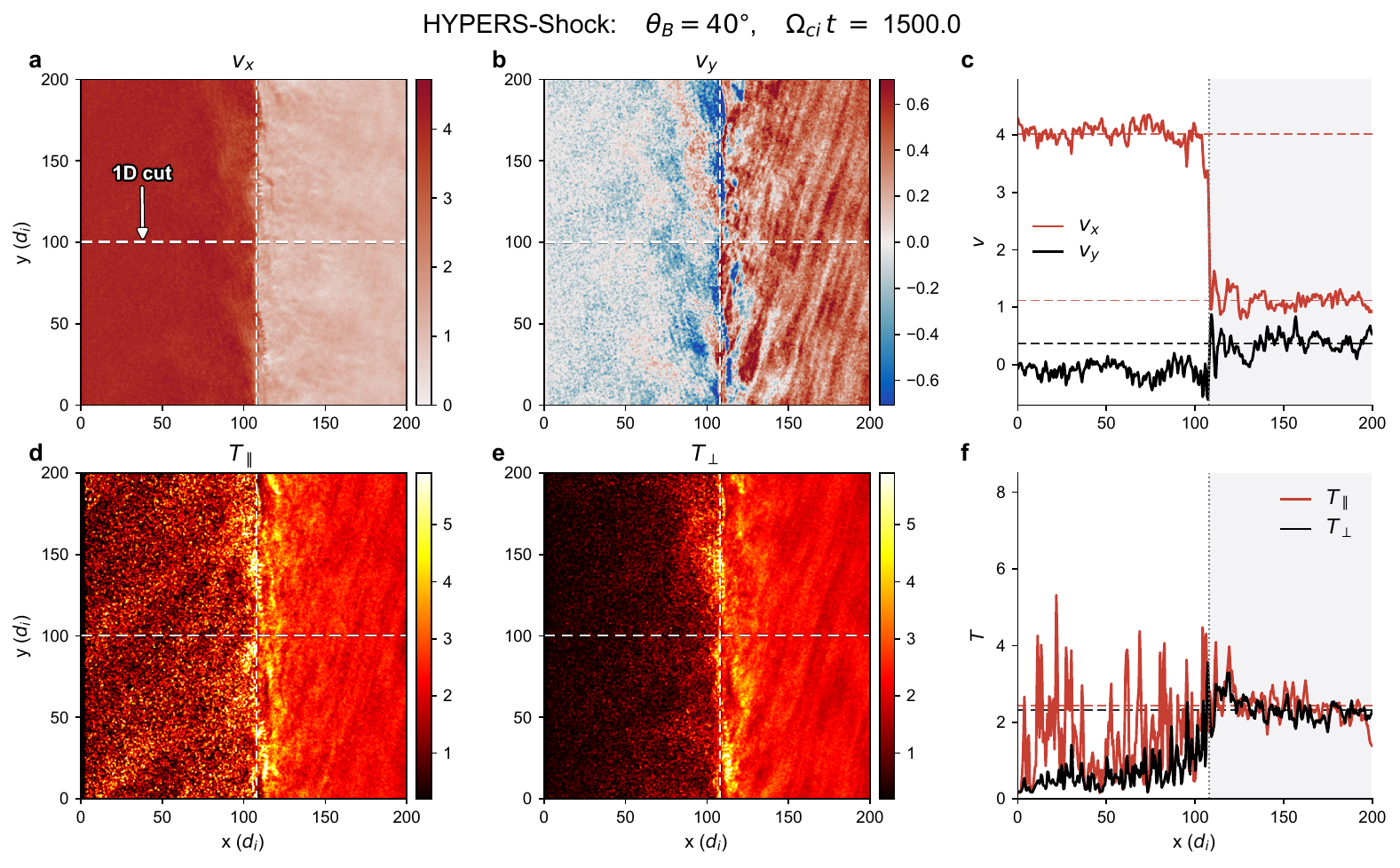}
\caption{Same as Figure~\ref{fig:mom20}, except for the oblique shock angle, $\theta_B=40^\circ$. The transverse flow velocity $v_y$ organizes into inclined, front-aligned shear bands in the downstream region, and the parallel and perpendicular downstream temperatures are comparable ($\langle T_\parallel\rangle\approx\langle T_\perp\rangle$; cf.\ Table~\ref{tab:runs}).}
\label{fig:mom40}
\end{figure}

\begin{figure}[p]
\centering
\includegraphics[width=\textwidth]{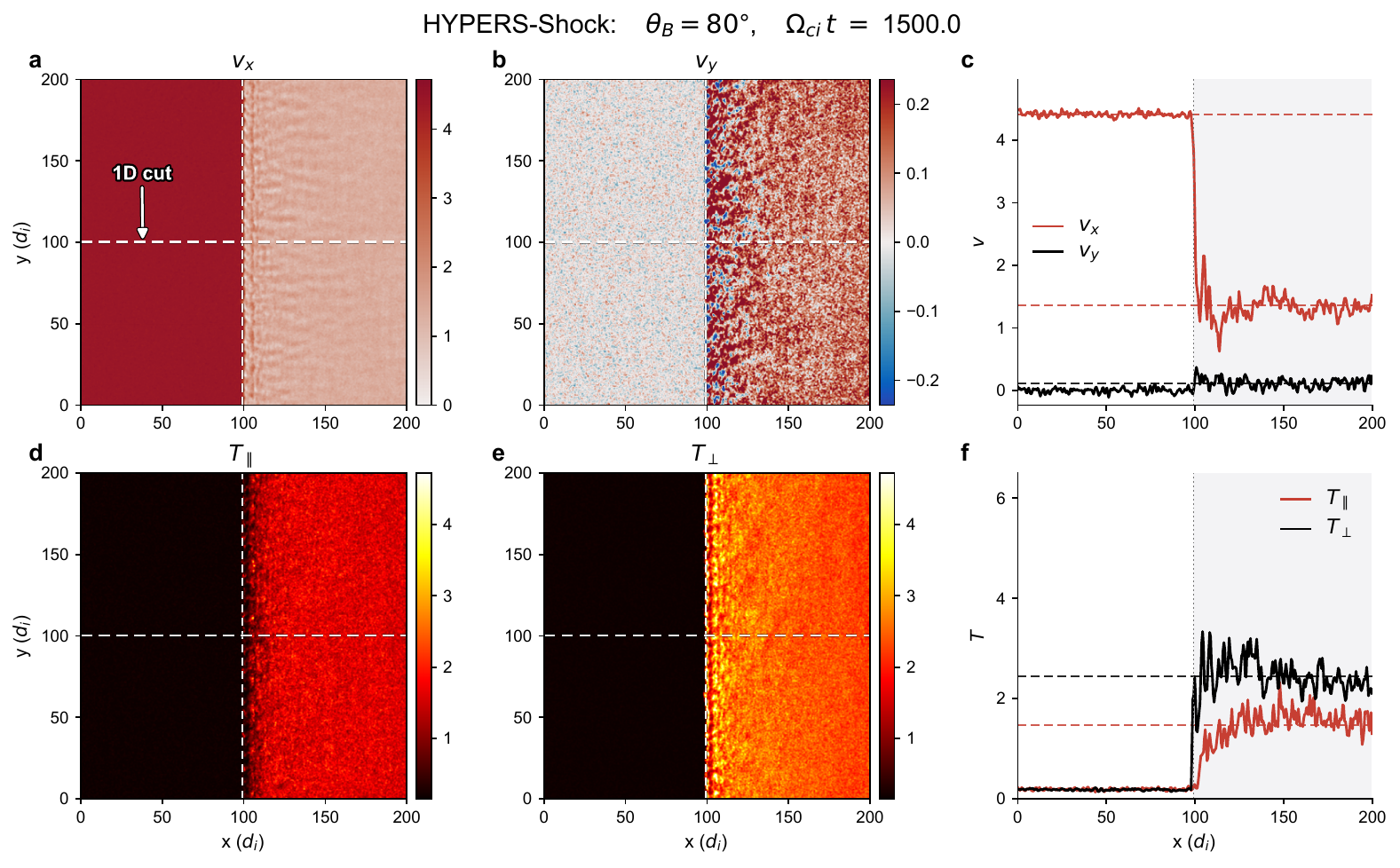}
\caption{Same as Figure~\ref{fig:mom20}, except for the quasi-perpendicular shock angle, $\theta_B=80^\circ$. The downstream transverse flow velocity $v_y$ is weak and turbulent, and the sharp shock front produces a downstream region in which the perpendicular heating dominates over the parallel heating, $\langle T_\perp\rangle>\langle T_\parallel\rangle$.}
\label{fig:mom80}
\end{figure}

Figures~\ref{fig:quad20}--\ref{fig:quad80} and 
Figures~\ref{fig:mom20}--\ref{fig:mom80} demonstrate 2D structures and 1D profiles of numerical shocks for three angles, $\theta_B=20^\circ, 40^\circ, 80^\circ$, selected to represent quasi-parallel, oblique, and quasi-perpendicular shocks, respectively. These simulation frames were recorded at the same representative time, $\Omega_{ci}t=1500$, chosen well past the transient period, $\Omega_{ci}t<750$. The asymptotic behavior of these shocks is consistent with time histories shown in Figures~\ref{fig:diagnostics}--\ref{fig:spectra}. As already mentioned, we created shock animations for all five values of $\theta_B$ \citep{Shockmovies2026}. These animations capture the initial plasma reflection off the right boundary, shock propagation to the middle of the computational domain, and subsequent controlled evolution of the quasi-stationary shock.

\textcolor{myred}{Figures~\ref{fig:quad20}--\ref{fig:quad80} illustrate a set of normalized simulation quantities: (a) electron density $n_e$; (b) magnetic-field magnitude $|\mathbf{B}|$, with the corresponding density and magnetic compression ratios, $r_n$ and $r_B$, respectively; (c) 1D cuts for $n_e$ (blue) and $|\mathbf{B}|$ (red) along the horizontal dashed line at $y=100\,d_i$ (see the arrow in panel~a), with the downstream plateau averages dashed and the downstream region shaded; (d,e) instantaneous local adaptive time increments for fields, $\Omega_{ci}\Delta t_F$ and particles, $\Omega_{ci}\Delta t_P$; and \textbf{(f)} 1D cuts for time increment distributions. The dashed vertical line marks the dynamic position of the shock front. The corresponding shock animations \citep{Shockmovies2026} clearly demonstrate plasma waves convecting from the foreshock region into the shock for different angles, $\theta_B$. We note that adaptive time updates enable accurate calculation of the transitional plasma features near the shock front. Earlier, adaptive temporal resolution was found to be critical for reproducing physically consistent plasma profiles in hybrid simulations of 1D collisionless shocks \citep{Omelchenko2006a}. 
}


\textcolor{myred}{It is important to note that for highly oblique shocks we see a significant growth of foreshock magnetic field oscillations within several ion gyro-periods after the ``piston-generated'' shock has been transformed into its co-moving frame (Figure~\ref{fig:frame_transform_n80}). This is an interesting finding in our study because high-obliquity moving shocks, generated with the ``piston method'', do not seem to have significant foreshock perturbations on their own (without imposing external turbulence on the upstream flow). Currently, we do not have a clear physical explanation for this computational phenomenon. It takes long times for fluctuations on the upstream side of moving oblique shocks to appear and grow \citep{preisser2020, kajdic2021}. We believe that larger-amplitude wave packets can form in the shock foot region even in highly oblique shocks (as confirmed by observations), but  ``piston-generated'' shocks are not simulated long enough to demonstrate their steady-stationary amplitudes. On the contrary, AFORA simulations achieve this foreshock state almost immediately after being transformed into the shock frame.
}

Figures~\ref{fig:mom20}--\ref{fig:mom80}, which accompany Figures~\ref{fig:quad20}--\ref{fig:quad80}, show ion bulk-flow and temperature moments. Across the full range of $\theta_B$ values used, the shock front, marked with a dashed vertical line and determined by locating the y-averaged density jump, clearly stands near the middle of the simulation box, $x_\mathrm{sh}\simeq100\,d_i$. The shock morphology, however, changes markedly with $\theta_B$. For a quasi-parallel shock, $\theta_B=20^\circ$, the shock front, shown in Figures~\ref{fig:quad20} and~\ref{fig:mom20}, is broad and strongly rippled, accompanied by an extended upstream region of large-amplitude compressive magnetic and whistler fluctuations generated by shock-reflected ions. The magnetic compression in this quasi-parallel case is modest, $r_B\simeq2.0$,  while the plasma density compression factor is close to the strong-shock limit ($r_n\simeq3.9$). This effect becomes even more pronounced for lower values of $\theta_B$, as seen in the $\theta_B=0$ shock animation \citep{Shockmovies2026}. As $\theta_B$ increases, the shock transition region sharpens and the magnetic field is increasingly compressed in step with the plasma density, until at a quasi-perpendicular angle, $\theta_B=80^\circ$ (see Figure~\ref{fig:quad80}) the shock front turns into a thin, nearly one-dimensional layer with the electron density, $n_e$, and magnetic field amplitude, $|\mathbf{B}|$, jumping across the shock front with close compression factors, $r_n\simeq r_B\simeq3.2$.

The bottom rows of Figures~\ref{fig:quad20}--\ref{fig:quad80} show that self-adaptive field ($\Omega_{ci}\Delta t_F$) and particle ($\Omega_{ci}\Delta t_P$) time increments contract significantly (by up to an order of magnitude for the fields) at the shock front, where plasma waves and particle orbits evolve fastest, compared to the relatively quiet upstream region. 

\textcolor{myred}{Figures~\ref{fig:mom20}--\ref{fig:mom80} illustrate the plasma flow deceleration, deflection, and heating during the shock transition. 
The upstream inflow velocity, $v_x=V_u'\simeq4\,V_A$, in the shock frame, decelerates abruptly at the front to the downstream value, consistent with the observed density compression ratio. The transverse flow velocity, $v_y$, develops in the downstream region so that it is strongest and most structured in the quasi-parallel and oblique shocks (Figures~\ref{fig:mom20}--\ref{fig:mom40}), being much weaker in the quasi-perpendicular shock (Figure~\ref{fig:mom80}).
}

Likewise, shock-driven plasma heating is $\theta_B$ dependent. In the upstream region, the quasi-parallel and oblique shocks (Figures~\ref{fig:mom20}--\ref{fig:mom40}) heat ions preferentially along the magnetic field. This effect clearly manifests itself in the form of field-aligned beams of hot backstreaming ions, extending far upstream. At the same time, the downstream plasma is isotropically thermalized, so that its perpendicular and parallel temperatures become roughly equal, $T_\perp\simeq T_\parallel$. By contrast, the quasi-perpendicular shock (Figure~\ref{fig:mom80}) produces a sharp temperature jump at the front, with the perpendicular temperature  exceeding the parallel one in the downstream region, where both temperatures reach distinctly different plateau values, $T_\perp > T_\parallel$.

Table~\ref{tab:runs} summarizes characteristic quasi-stationary shock quantities as a function of shock obliquity angle, $\theta_B$. These quantities, time-averaged over a subset of simulation shots, $\Omega_{ci}t\geq750$, are as follows: the upstream and downstream flow speeds, magnetic and density compression ratios, downstream parallel and perpendicular temperatures, and ion-acceleration efficiency metrics. \textcolor{myred}{To produce these time-averaged upstream and downstream characteristics, we first obtain single values for each time frame. The shock front for these purposes is identified as the steepest rise of the transverse-averaged density profile. A single-value frame quantity is then obtained by averaging first along the transverse y-direction, followed by averaging over the upstream or downstream plateaus, respectively. 
}

\textcolor{myred}{\subsection{Collisionless Shocks in Hybrid Simulations and Observations}
}

\textcolor{myred}{Magnetic field profiles of shocks in self-consistent hybrid simulations often appear to lack significant perturbations in the upstream (foot) and downstream regions, especially for oblique shocks \citep[e.g.,][]{caprioli2014,gedalin2018, preisser2020, trotta2023, gedalin2023}.
Several factors may contribute to causing excessive numerical diffusion in hybrid simulations \citep{Shay2025Simulation}, making shock profiles appear overly smooth. For instance, {\it dHybrid} simulations do not maintain a numerical divergence-free magnetic field \citep{Caprioli2018} and use ad hoc filtering to remove short wavelengths \citep{gedalin2018}. On the other hand, the HYPSI \citep{preisser2020} or H3D  \citep{gedalin2023} codes apply large numerical resistivity to suppress unphysical (short-wavelength) whistler modes. 
}

\textcolor{myred}{Collisionless shocks, regularly observed near Earth, are known to exhibit coherent downstream oscillations, as well as enhanced wave activity in the upstream and downstream regions of the shock and near its ramp \citep{gedalin2018, gedalin2022, gedalin2026}. Examples of MMS measurements of magnetic field and density profiles of space shocks are shown in Figure~\ref{fig:observations}. This Figure compares the two selected shock
crossings. The first is an Earth's bow shock event observed by MMS on 2017
October 12 at approximately 15:22:29 UT, with
$\theta_{Bn}\simeq59^\circ$. The second is an interplanetary shock event observed by MMS3 on 2018 January 8 at approximately 06:41:10 UT, with
$\theta_{Bn}\simeq66^\circ$. The bow shock is therefore somewhat more
oblique, while the interplanetary event lies closer to the
quasi-perpendicular regime.}

\textcolor{myred}{
These two events show well-defined transitions in the magnetic field magnitude and electron density. There is a good qualitative agreement between these observational profiles and the magnetic field and density profiles of our numerical shocks (e.g., Figures~\ref{fig:quad20}--\ref{fig:quad80}, panel (c), and temporal dynamics in \citealt{Shockmovies2026}). This analysis includes comparing such important shock characteristics as the large-amplitude upstream waves in the shock foot, the sharp ramp, the magnetic overshoot, and the quasi-periodic oscillations in the downstream region (for more detail see \citealt{gedalin2026}). 
}

\textcolor{myred}{As noted above, it is difficult to clearly separate essential physical effects and transient numerical features in the structure of a moving shock produced by the ``piston'' method \citep{bennett1995}. Because of this physical uncertainty in moving shock simulations, test-particle simulations \citep{gedalin2018, hanson2020} have been used to study ion dynamics and related magnetic perturbations in the shock frame. For low-Mach number shocks, test-particle simulations by \citet{gedalin2018} were found to produce magnetic profiles with downstream perturbations similar to observed ones. At the same time, self-consistent hybrid simulations, conducted in the same study with similar parameters, failed to reproduce the upstream waves seen in shock observations. The latter was attributed to numerical smoothing effects that effectively damped out upstream oscillations in the hybrid model.
}

\textcolor{myred}{
Adequate resolution of upstream and downstream magnetic perturbations in self-consistent hybrid simulations is crucial for properly modeling both the DSA and SDA mechanisms of ion acceleration.  In SDA, wave effects may affect the dynamics of ions traveling along the shock front and gyrating through the ramp before they either pass into the downstream or escape into the upstream region. Accordingly, the efficiency of DSA critically depends on the time spent by ions within the strong magnetic field gradients of the shock transition \citep{hanson2020}.
}

\begin{table}[!ht]
\centering
\caption{Characteristic shock quantities for all values of $\theta_B$ used in the simulations. Each entry is an average over all simulation shots taken during the time period, $\Omega_{ci}t\ge750$, i.e., beyond the cutoff moment, after which shock behavior becomes quasi-stationary. 
The percentage value in parentheses is the relative standard deviation (scatter error) with respect to the mean;
it reads ``N/A'' when the mean value is close to zero. 
$\langle v_x\rangle_\mathrm{up}$ and $\langle v_y\rangle_\mathrm{down}$ are the upstream averaged $v_x$ and downstream averaged $v_y$ components of the plasma velocity, respectively; $\langle T_\parallel\rangle$ and $\langle T_\perp\rangle$ are the downstream averaged parallel and perpendicular temperatures; $r_B$ and $r_n$ are the plateau-averaged downstream to upstream compression ratios, computed with respect to the absolute value of magnetic field, $|\mathbf{B}|$, and plasma density, $n_e$, respectively. Ion energies, $E$, are normalized to the upstream injection energy, $E_0$. The cumulative efficiency of downstream ion acceleration is represented by two integrals of ion distributions over $E\geq 1$ (see Figure~\ref{fig:spectra}): $\langle\mathrm{d}n_i\rangle_t=\int_{E\ge1}\langle\mathrm{d}n_i/\mathrm{d}E\rangle_t\,\mathrm{d}E$ and $\langle\mathrm{d}n_iE\rangle_t=\int_{E\ge1}\langle\mathrm{d}n_i/\mathrm{d}E\rangle_t\,E\,\mathrm{d}E$, representing the particle number and energy densities, respectively.}
\label{tab:runs}
\small
\setlength{\tabcolsep}{4.5pt}
\begin{tabular}{c cc cc cc cc}
\toprule
$\theta_B$ & \multicolumn{2}{c}{Flow} & \multicolumn{2}{c}{Compression} & \multicolumn{2}{c}{Downstream $T$} & \multicolumn{2}{c}{Efficiency} \\
\cmidrule(lr){2-3}\cmidrule(lr){4-5}\cmidrule(lr){6-7}\cmidrule(lr){8-9}
(deg) & $\langle v_x\rangle_\mathrm{up}$ & $\langle v_y\rangle_\mathrm{down}$ & $r_B$ & $r_n$ & $\langle T_\parallel\rangle$ & $\langle T_\perp\rangle$ & $\langle\mathrm{d}n_i\rangle_t$ & $\langle\mathrm{d}n_iE\rangle_t$ \\
\midrule
0  & 4.1 (0.3\%) & 0.0 (N/A)   & 1.7 (5.7\%) & 3.6 (1.6\%) & 2.8 (3.5\%) & 2.7 (2.5\%) & 1.2 (3.7\%) & 2.0 (4.2\%) \\
20 & 4.0 (0.4\%) & 0.2 (7.1\%) & 1.8 (2.0\%) & 3.7 (1.9\%) & 2.8 (2.5\%) & 2.6 (2.9\%) & 1.4 (6.2\%) & 2.3 (7.6\%) \\
40 & 4.0 (0.3\%) & 0.4 (2.6\%) & 2.7 (1.1\%) & 3.5 (1.2\%) & 2.4 (1.7\%) & 2.3 (2.0\%) & 1.1 (3.8\%) & 2.0 (4.1\%) \\
60 & 4.4 (0.1\%) & 0.3 (1.5\%) & 2.9 (0.5\%) & 3.2 (0.5\%) & 2.5 (1.4\%) & 2.7 (1.4\%) & 0.8 (2.8\%) & 1.5 (2.8\%) \\
80 & 4.4 (0.7\%) & 0.1 (3.9\%) & 3.1 (2.0\%) & 3.1 (2.0\%) & 1.6 (6.7\%) & 2.4 (4.5\%) & 0.7 (7.4\%) & 1.2 (8.4\%) \\
\bottomrule
\end{tabular}
\end{table}

\begin{figure}[!ht]
\centering
\includegraphics[width=\textwidth]{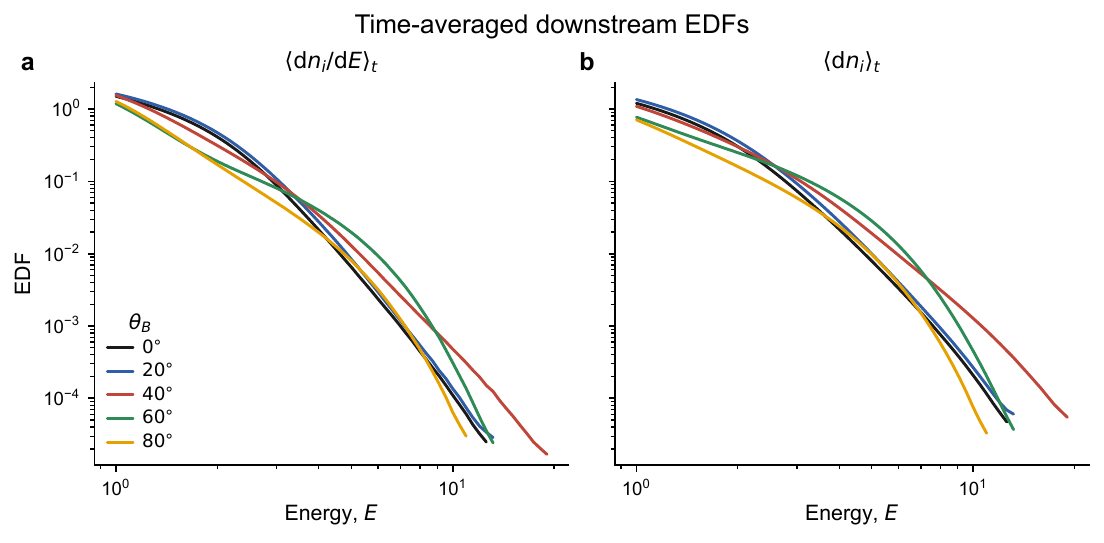}
\caption{Time-averaged downstream ion energy distributions for all five values of the magnetic inclination angle, $\theta_B$, used in this work: \textbf{(a)} the differential spectra, $\langle\mathrm{d}n_i/\mathrm{d}E\rangle_t$, and \textbf{(b)} the cumulative spectra, $\langle\mathrm{d}n_i\rangle_t=\int_{E'\ge E}\langle\mathrm{d}n_i/\mathrm{d}E'\rangle_t\,\mathrm{d}E'$. One solid curve per angle is used, colored as in Figure~\ref{fig:diagnostics}; both panels share the same scales, and ion energies are normalized to the injection energy, $E_0$, as in Table~\ref{tab:runs}. Each curve averages the 111 shots recorded past the cutoff time, $\Omega_{ci}t\ge750$. For $\theta_B=40^\circ$ and $60^\circ$, these are the same averages, drawn as black curves in Figure~\ref{fig:spectra}. Panel (b) values, taken at $E=E_0$, reproduce the $\langle\mathrm{d}n_i\rangle_t$ column of Table~\ref{tab:runs}. Energy bins, recording values below $10^{-5}$, hold too few ions to be statistically meaningful and are excluded from these distributions; each curve is drawn up to the highest energy at which at least 44 of its 111 shots remain populated, so that the curve endpoint effectively marks the highest energy, where the tail ceases to be persistently sampled in time.
}
\label{fig:spectra_angles}
\end{figure}

\textcolor{myred}{\subsection{Post-Shock Acceleration of Seed Ions in the Downstream Region}
}

\textcolor{myred}{The HYPERS-Shock model implements a run-time diagnostic tool for computing energy spectra of accelerated ions. Ion energies are computed in the solar wind frame, above the ion injection energy, $E_0$, used for energy normalization (see Section~\ref{sec:HYPERS}). Unlike \textit{a posteriori} particle diagnostics, which normally process limited subsets of particles randomly recorded during the simulation, the HYPERS-Shock routine enables parallel, accurate, and adaptive runtime computation of ion energy distributions in any given subdomain (box). 
}

\textcolor{myred}{In this paper, we focus on ion spectra obtained in the downstream regions of shocks of different obliquities. Physically, these downstream spectra represent energy distributions of post-shock ions \citep{caprioli2014} that escape upstream, where they continue to be accelerated by the upstream turbulence to much higher energies, producing SEPs at Earth's orbit \citep{sokolov2004new} and CRs at astrophysical shocks \citep{simon2026}. We emphasize that our main goal is to investigate the quasi-stationarity of these spectra in the shock frame and provide a proof that these spectra remain quasi-steady for long simulation times. Given this proof (see below), we can safely generate these spectra at much shorter simulation times, as needed. In this study, we use the same Alfv\'en Mach number, $M_A=3$, and the same ratio, $V_A/c$, in the downstream frame (see Section~\ref{sec:HYPERS}), to investigate the dependence of ion spectra on shock obliquity only. In the future, we plan to study the efficiency of ``seed'' ion acceleration for heliospheric shocks in physical energy units (keV) as a function of these two parameters. We also note that downstream ion acceleration has been previously investigated by \citet{caprioli2014} for moving shocks with $M_A\geq5$, who also used different (from ours) metrics to represent acceleration efficiency. They also needed much longer domains to enable convergence of ion spectra generated by very energetic shocks. A direct comparison of our spectra to theirs is impossible in the framework of this study, but it may be done in the future.
}

The ion spectra in our simulations were found not to be sensitive to the length of their downstream recording box in the x-direction, up to $0.3$ times the length of the simulation domain. Presently, we do not resort to increasing the resolution of energy spectra through particle splitting \citep[see, e.g.,][]{bennett1995}. This feature is, however, implemented in the HYPERS-Shock model and can be used for modeling more energetic spectra in the future. 

We estimate the efficiency of downstream ion acceleration in the solar wind frame by computing time-averaged ion number and energy densities, as well as ion differential and cumulative energy distribution functions for all values of $\theta_B$ used in our simulations (see Table~\ref{tab:runs}). Figure~\ref{fig:spectra} illustrates these spectral data for $\theta_B=40^\circ$ and $\theta_B=60^\circ$. These two angles straddle the $45^\circ$ boundary between the quasi-parallel and quasi-perpendicular regimes. Time-averaged spectra obtained for all five angles are collected in Figure~\ref{fig:spectra_angles}, which compares them directly on common scales. The time-averaged cumulative energetic ion number density, $\langle\mathrm{d}n_i\rangle_t$ (panel (a)), can be used as input for Fokker-Planck simulations of ion acceleration by solar wind turbulence \citep{sokolov2004new}.

Two other diagnostics in Figure~\ref{fig:spectra} serve several purposes. They (i) compute the time-averaged differential number density of ions as a function of energy, $\langle\mathrm{d}n_i/\mathrm{d}E \rangle_t$ (panel (a)), and (ii) prove the fast convergence of ion spectra in the shock frame (panels (b) and (c)). The latter metrics, illustrated in panels (b) and (c), clearly demonstrate that the downstream distributions of energetic ions remain quasi-stationary after the cutoff time, $\Omega_{ci}t>750$. Individual differential density spectra, $dn_i/dE$, recorded after this cutoff, were used to compute the time-averaged spectrum $\langle dn_i/dE\rangle_t$ and the corresponding cumulative spectrum $\langle dn_i \rangle_t$ (panel~a). The shot-to-shot relative scatter error in these calculations stays at a few-percent level across well-sampled energies and rises only in the sparsely populated high-energy tail (panel~b). This error decreases and then plateaus/drops as the early frames ($\Omega_{ci}t<750$, not shown in panel (a)) are excluded from calculations.
\textcolor{myred}{As follows from Table~\ref{tab:runs}, the efficiency of downstream ion acceleration has been found to be greater for quasi-parallel shocks (smaller $\theta_B$), consistent with the DSA mechanism and findings by \citet{caprioli2014}. However, in our simulations, this efficiency does not drop dramatically for larger values of $\theta_B$ either, with energetic ion tails reaching energies $E>10E_0$ for all angles considered in this work.
}

In Figure~\ref{fig:spectra_angles}, we plotted together time-averaged energy distribution functions for all five values of $\theta_B$ used in this work. The comparative analysis of these spectra shows that the number of seed ions and the extent of their high-energy tail vary with  $\theta_B$, being inversely proportional to each other. 
Below $E\simeq3.3E_0$, quasi-parallel shocks carry larger ion populations ($\theta_B=0^\circ$ and $\theta_B=20^\circ$ are the highest curves), with the seed ion density, $\langle\mathrm{d}n_i\rangle_t$ exceeding the corresponding value for  $\theta_B=80^\circ$ by a factor of up to $\simeq2.8$ near $E=2E_0$. This energy interval holds $\gtrsim 85\%$--$94\%$
of all the accelerated ions and therefore sets the ordering of the corresponding acceleration efficiency columns in Table~\ref{tab:runs}.
Above $E\simeq3.3E_0$, this efficiency ordering reverses and oblique shocks ($\theta_B=40^\circ$ and $\theta_B=60^\circ$) lead in efficiency, by a factor of up to $\simeq3.6$. Of the five angles used, we find that $\theta_B=40^\circ$ is the hardest in slope ($-4.8$ fitted over $5E_0\le E\le10E_0$ against $\simeq-5.9$ at $\theta_B=0^\circ, 20^\circ$, and $60^\circ$ and $-6.9$ at $80^\circ$), having the tail extending the farthest, to $E\gtrsim19E_0$. Every angle, nevertheless, sustains a persistently sampled tail, $E\geq 10E_0$, even with the quasi-perpendicular curve ending at $E\simeq11E_0$. The most prominent high-energy tails are seen at intermediate (oblique) angles, $\theta_B=40^\circ,60^\circ$.

\section{Summary and Conclusions}
\label{sec:summ}

\textcolor{myred}{The short-timescale energy dissipation processes in heliospheric and astrophysical shocks, which lead to acceleration and injection of ions into the solar wind for their further energization, strongly depend on the shock structure and require kinetic physics. Quasi-neutral hybrid simulations (kinetic ions, fluid electrons) are very well suited for modeling these processes. Hybrid simulations of moving shocks, however, are intrinsically time dependent, and in addition require long simulation domains for studying upstream acceleration of ions (such as cosmic rays). As a result, these models remain very demanding of computational resources even in two dimensions.
}

\textcolor{myred}{In this paper, we have presented a novel computational technique (AFORA) for modeling quasi-steady collisionless plasma shocks with arbitrary angles between the ambient magnetic field and the shock normal (shock obliquities). This computational approach has been shown to generate a steady-state self-consistent shock structure and concomitant quasi-stationary downstream ion spectra in a computational domain with minimum dimensions. It can be applied to study the physics of energetic ions either downstream (``seed ions'') or upstream (``cosmic rays'') of the shock. We have demonstrated this method for quasi-parallel, oblique, and quasi-perpendicular shocks and generated movies that visualize quasi-steady dynamics of these shocks in the shock frame \citep{Shockmovies2026}. 
}

The AFORA technique has been implemented in a multiscale hybrid code, HYPERS, routinely used to conduct global simulations of solar wind interactions with Earth's magnetosphere in two and three dimensions \citep{Omelchenko2012a, Omelchenko2023spr}. HYPERS offers important advantages over standard hybrid models in the way electromagnetic fields and particles are advanced in time. Instead of specifying global timesteps, the HYPERS operating system (EMAPS) adaptively updates computational cells and particles with their individual time increments, selected to match local physical rates of change. This event-driven simulation technology, previously used in 1D hybrid simulations to study plasma shocks and discontinuities \citep{Omelchenko2006a}, offers superior physical accuracy and computational efficiency compared to its time-stepping counterparts. In this paper, we have shown that the combination of AFORA and EMAPS in HYPERS expeditiously produces numerical shocks with profiles similar to those measured in interplanetary shocks commonly observed by spacecraft. 

\textcolor{myred}{Our approach offers two important advantages over simulations of moving shocks. First, any reliable analysis of moving shocks requires long simulation times because the shock constantly undergoes transitory changes, with upstream waves slowly growing at oblique shocks. 
Recently, \citet{simon2026} have developed a ``faux-shock'' simulation model for modeling cosmic rays generated upstream of astrophysical shocks. Their approach aims at reducing the length of the upstream region (to increase the computational efficiency of hybrid simulations) by enabling simulations in the shock frame. The ``faux'' shock is an artificial semi-transparent boundary with no self-consistent downstream region. In comparison, our approach enables a self-consistent hybrid simulation both upstream and downstream of the shock. Therefore, our shock-frame model (AFORA) should provide better physical fidelity when applied to the same class of astrophysical problems.
}

\textcolor{myred}{In the context of modeling ``seed ions'' (this setup is also known as the ``injection problem'' for heliospheric and astrophysical shocks), our method produces numerical shocks that reproduce many features of interplanetary (IP) shocks, commonly observed by spacecraft. In addition, AFORA offers fast and reliable convergence of quasi-stationary shock features and downstream ion spectra. 
Contrary to moving shocks, plasma shock simulations in the shock frame also enable efficient parallelization strategies and simplify mesh refinement around the shock front. For better spatial resolution and parallel load balancing of shock-frame simulations, we plan to enhance the HYPERS-Shock model with a static stretched mesh and matching stretched processor grid, already used in HYPERS simulations of Earth's bow shock and magnetosheath \citep{Omelchenko2021a}. 
}

\textcolor{myred}{In the past, 2D hybrid simulations of collisionless plasma shocks reproduced many physical features, critical to understanding the DSA and SDA mechanisms dominating ion acceleration in quasi-parallel and quasi-perpendicular shocks, respectively. More recently, however, \citet{orusa2023}, \citet{orusa2025}, and \citet{Caprioli2025} have pointed out that the efficiency of ion acceleration at perpendicular shocks may critically depend on the ability of simulation ions to scatter across the magnetic field efficiently, so that they can return from the downstream to the upstream and continue to gain energy at each bouncing cycle through the SDA mechanism. This ion motion can be fully enabled in 3D simulations only, which are significantly more CPU-intensive compared to similar 2D simulations. Using the AFORA methodology, we demonstrate significant levels of magnetic fluctuations in the foreshock (foot) region of quasi-perpendicular shocks modeled in the shock frame.  Magnetic field profiles of our shocks in the downstream region also seem to be more realistic than smoothed profiles found in other hybrid simulations. Together with these findings, future simulations of 3D shocks in the shock frame are expected to provide more accurate estimates of the efficiency of acceleration of ``seed'' ions at collisionless shocks of different obliquities. 
}

Finally, we plan to interface the HYPERS-Shock model with the M-FLAMPA (Multiple Field-Line-Advection Model for Particle Acceleration) code. M-FLAMPA models long-time acceleration and transport of SEPs in the background solar wind turbulence along continuously evolving solar magnetic field lines, where the magnitudes of electromagnetic fields are parameterized using available observations. As input, M-FLAMPA accepts densities of seed energetic ion populations, accelerated at heliospheric (CME-driven) shocks to energies above $\sim 10~\mathrm{keV}$. 

\section{Data Availability and Support Statement}
The authors were partially supported by NASA Award 80NSSC23M0191 (The CLEAR Space Weather Center of Excellence). Yuri Omelchenko was also partially supported by NSF-BSF Award 2512084 and NASA Award 80NSSC26K0056. The computing resources supporting this work were provided by the NASA High-End Computing (HEC) Program through the NASA Advanced Supercomputing (NAS) Division at the Ames Research Center. The reduced dataset, simulation files, and post-processing scripts, used to present results from this research, are openly available in Zenodo at https://doi.org/10.5281/zenodo.21286632.

\clearpage
\bibliographystyle{aasjournal}

\end{document}